\documentclass{ws-procs9x6}

\newcommand{\nn}{\nonumber}
\newcommand{\be}{\begin{equation}}
\newcommand{\ee}{\end{equation}}
\newcommand{\ba}{\begin{eqnarray}}
\newcommand{\ea}{\end{eqnarray}}
\newcommand{\req}[1]{(\ref{#1})}

\def\gev{~{\rm GeV}}
\def\ale{\alpha_{\rm elm}}
\def\als{\alpha_{\rm s}}

\def\qbq{q\overline{q}}
\newcommand{\lsim}{\raisebox{-4pt}{$\,\stackrel{\textstyle
                                                         <}{\sim}\,$}}

\newcommand{\ov}[1]{\overline#1} 
\newcommand{\sla}{\hspace*{-0.20cm}/}
\newcommand{\da}{{distribution amplitude}}

\begin{document}
\title{Spin and hard processes}

\author{P.\ Kroll}

\address{Fachbereich Physik, Universit\"at Wuppertal,\\ 
D-42097 Wuppertal, Germany\\
Email: kroll@physik.uni-wuppertal.de}
\maketitle
\abstracts{It is argued that spin is a fundamental aspect of gauge
theories at short distances. As a consequence there are
characteristic helicity asymmtries in hard inclusive and exclusive
reactions  of which a few are discussed.}  
\section{Introduction}
The jibe, occasionally heard in the late sixties and early seventies, that
spin is an inessential complication of elementary particle physics,
does not match reality. On the contrary, spin is a fundamental aspect
of gauge theories at short distances. In the electroweak theory, based
on broken SU(2)$\times$U(1) gauge symmetry, the left-handed quarks and
leptons form SU(2) doublets, the right-handed ones singlets while
right-handed neutrinos do not exist. These characteristics evidently
lead to a wealth of polarization phenomena. Although it is not so
obvious, QCD leads to characteristic spin dependences, too. The basis
of any calculation within QCD is the factorization of a reaction into
a hard parton-level subprocess to be calculated from perturbative
QCD and/or QED, and process-independent soft hadronic matrix elements 
which are subject to non-perturbative QCD and are not calculable  
to a sufficient degree of accuracy at present. Factorization has been
shown to hold for a number of inclusive and exclusive reactions
provided a large momentum scale (corresponding to short distances) is
present. For other reactions 
factorization is a reasonable hypothesis as yet. In the absence of a
large scale we do not know how to apply QCD and, for the
interpretation of scattering reactions, we have to rely upon effective 
theories or phenomenological models as for instance the Regge pole one.  

If an almost massless quark interacts with a number of gluons and/or
photons, helicity flips are suppressed since 
\be
\ov{u}(-)\gamma^\mu (q\sla{}_1 \gamma^{\mu_1} \cdots q\sla{}_n \gamma^{\mu_n})
                             u(+) \propto \frac{m_q}{Q}\,.
\label{eq:flip}
\ee
The current quark mass, $m_q$, is of order MeV while the hard scale,
$Q$, is of order GeV. Therefore, to a very good approximation, a quark
line will always carry the same helicity, i.e.\ quark helicity is
conserved. Now, to leading-twist order which dominates in hard processes, 
the helicity of the quark is transferred to its parent
hadron to a large extent. From these considerations follows that {\it
helicity asymmetries of hadronic processes reflect the route of the
quark lines through the process} \cite{jacob83}.
\section{Inclusive reactions}
\begin{figure}[t]
\begin{center}
\includegraphics[width=8.0cm]{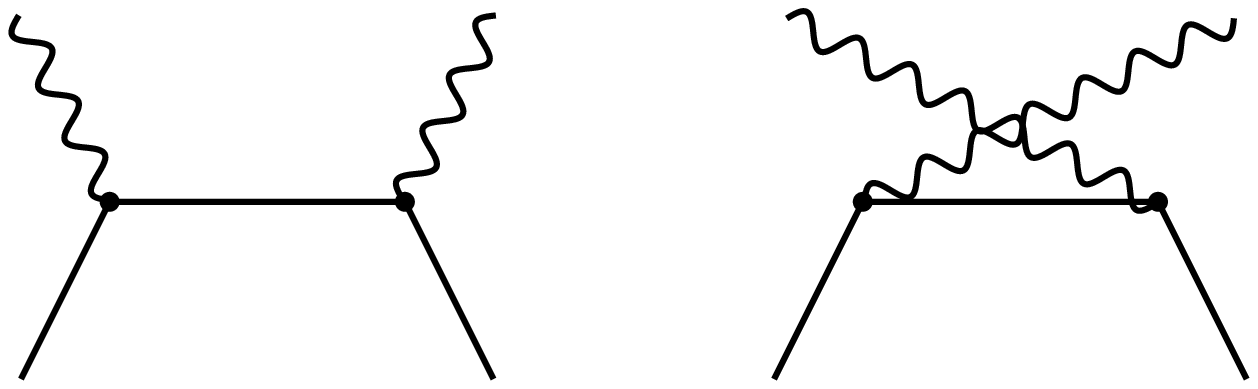}   
\caption{Lowest order Feynman graphs for Compton scattering off
quarks. The wavy lines represent either gluons or photons.}
\label{fig:csquark}
\end{center}
\end{figure}

As an example let me discuss prompt photoproduction, $A B\to
\gamma X$, at large transverse momentum of the produced photon. Two
parton-level subprocesses contribute, $\qbq\to \gamma g$ and $gq\to
\gamma q$. In the first process quark and antiquark have opposite
helicities according to \req{eq:flip} leading to the parton-level
helicity correlation 
\be
\hat{A}_{LL}(\qbq\to \gamma g)= \frac{d\hat{\sigma}(\lambda_q=+, \lambda_{\ov{q}}=+) 
                   -d\hat{\sigma}(\lambda_q=+, \lambda_{\ov{q}}=-)}
                    {d\hat{\sigma}(\lambda_q=+, \lambda_{\ov{q}}=+) 
                       +d\hat{\sigma}(\lambda_q=+,
		       \lambda_{\ov{q}}=-)} = -1\,.
\ee
In the second process, the correlation of the incoming gluon and 
quark helicities reads (the relevant Feynman graphs are shown in Fig.\
\ref{fig:csquark})
\be
\hat{A}_{LL}(gq\to \gamma q)= \frac{\hat{s}^2-\hat{u}^2}
{\hat{s}^2+\hat{u}^2}\,,  
\label{eq:allhat}
\ee
where $\hat{s}$ and $\hat{u}$ are the subprocess Mandelstam variables.
We note that \req{eq:allhat} holds for $gq\to gq$ and $\gamma q\to
\gamma q$ as well. The subprocess helicity correlations lead to
characteristic differences in the proton-antiproton helicity
correlations of, say, $p\ov{p}\to\gamma X$, where the $\qbq$ subprocess 
dominates, and $pp\to\gamma X$ which is under control of 
$gq\to \gamma q$ \cite{Vogelsang93}. 
\section{Leading-twist factorization in hard exclusive processes}
\begin{figure}[t]
\begin{center}
\includegraphics[width=4.8cm,bbllx=73pt,bblly=525pt,bburx=530pt,
bbury=795pt,clip=true]{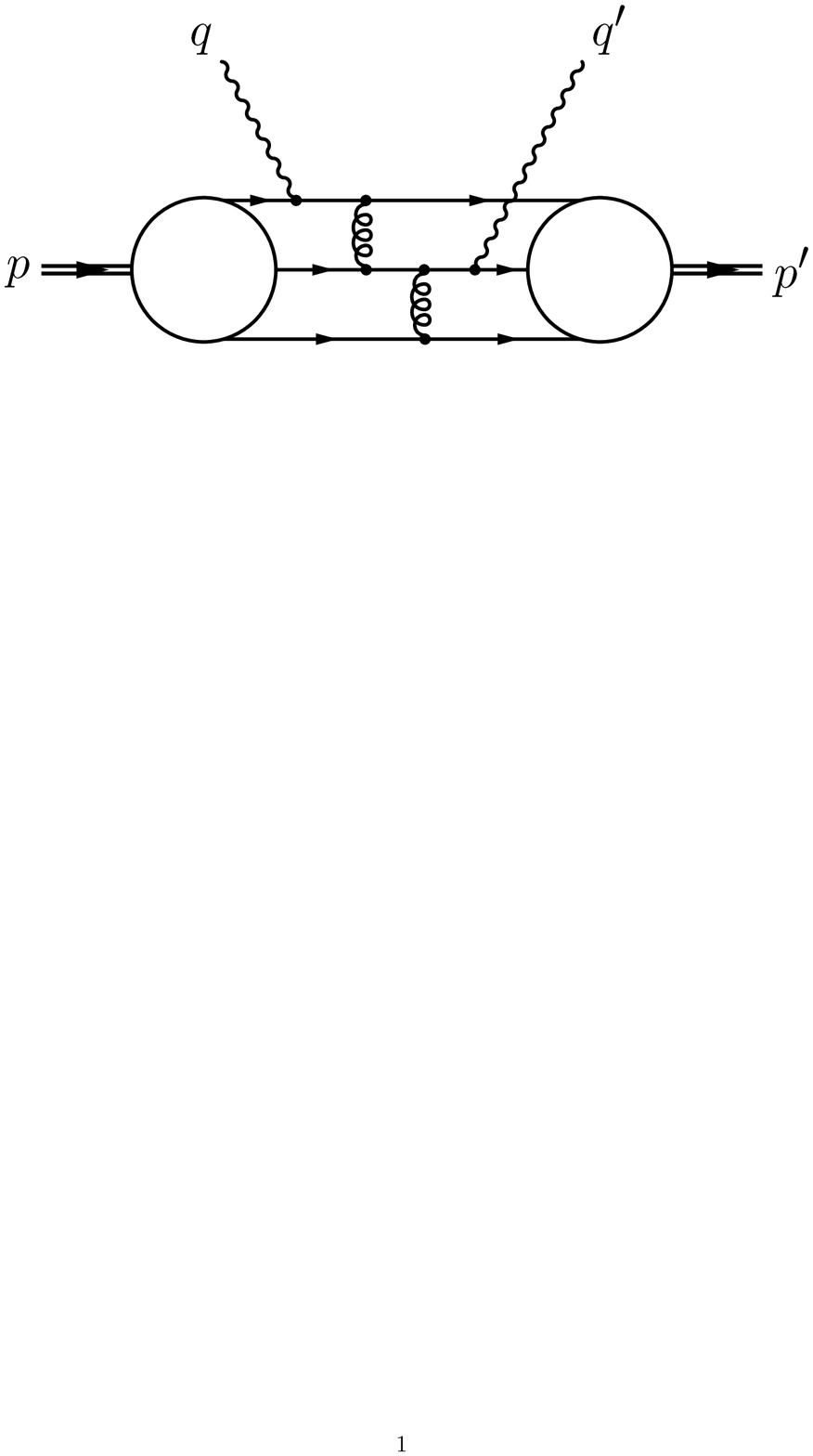} \hspace*{0.5cm} 
\includegraphics[width=3.2cm,bbllx=55pt,bblly=230pt,bburx=545pt,
bbury=615pt,clip=true]{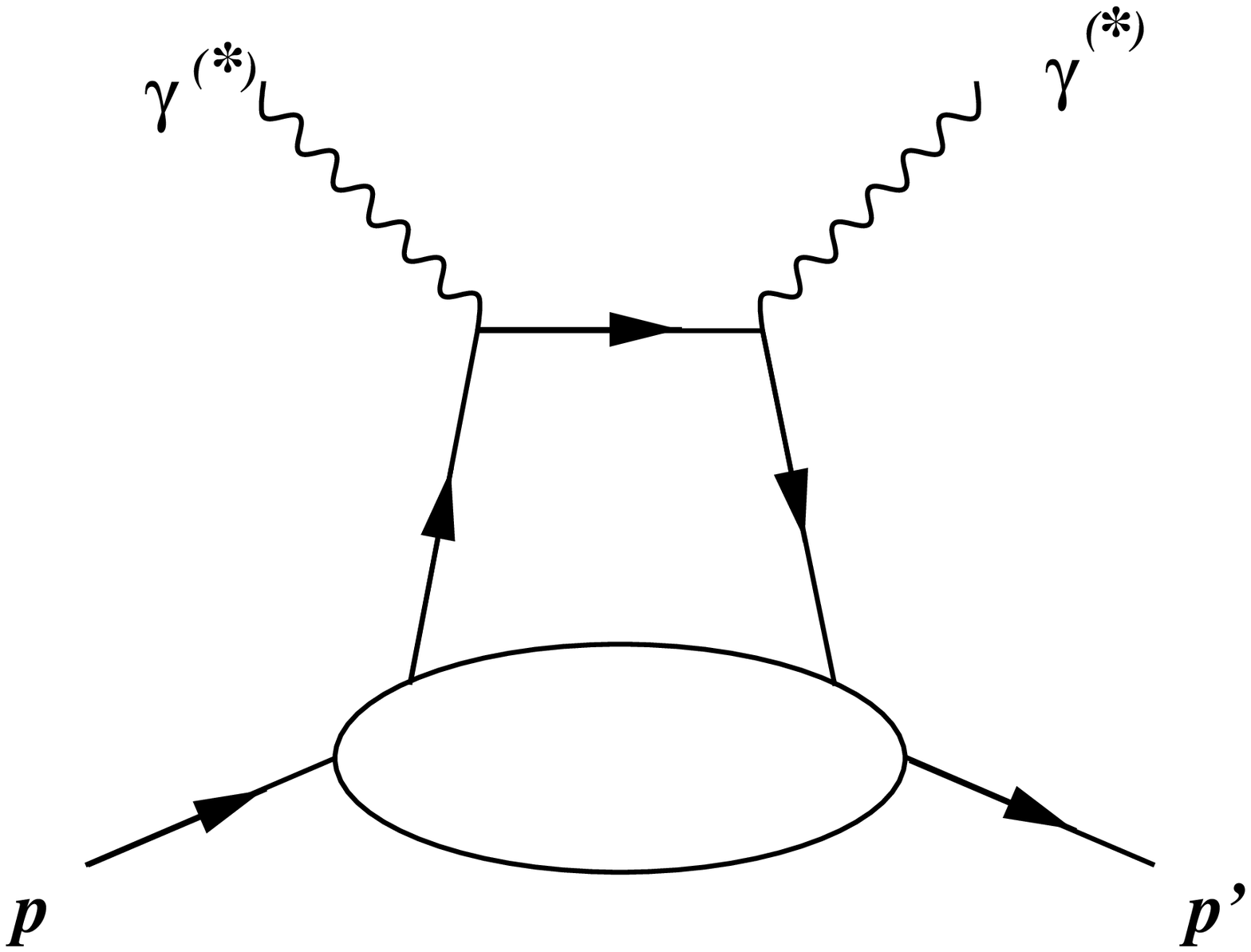} 
\caption{{}Leading-twist (left) and handbag (right) factorization for
Compton scattering.}
\label{fig:handbag}
\end{center}
\end{figure}

For asymptotically large $s, -t, -u$ the dominant (leading-twist)
contribution to an exclusive reaction is produced by the valence quarks 
of the involved hadrons \cite{bro80}. The quarks move approximately 
collinear with their parent hadrons and participate in the hard
scattering while the soft physics is encoded in \da s, $\Phi(x_1,
\dots x_n)$, representing the momentum distribution of the quarks in a
hadron. For Compton scattering, for instance, the hard process is
$\gamma qqq\to \gamma qqq$, see Fig.\ \ref{fig:handbag}, and the
Compton amplitude is given by the convolution  
\be 
M = \Phi \otimes  H \otimes \Phi \,,
\label{eq:conv}
\ee
where ${H}$ is the parton-level subprocess amplitude.

To leading-twist accuracy the helicities of the valence quarks,
conserved in the hard process, sum up to the parent's hadron helicity,
there is no quark orbital angular momentum, $L_q$,
involved. Configurations where the hadron helicity differs from
the sum of the valence quark helicities which obviously require
$L_q\neq 0$, are of higher-twist nature and are suppressed by inverse
powers of the hard scale, $-t(-u)$, as compared to the leading-twist 
contribution. Hence, to leading-twist order, the conservation of quark
helicity converts into hadronic helicity conservation for all hadrons
that are connected by light quark lines. 
Experimentally however hadronic helicity conservation is violated for
hard scales of the order 10 GeV$^2$; the ratio fo flip to non-flip
amplitudes is typically $20-30\%$. Examples for reactions where such
violations have been observed are the Pauli form factor of the
proton~\cite{gayou}, the polarization in proton-proton elastic
scattering~\cite{krisch} or the charmonium decays
$\eta_c(\chi_{c0})\to p\ov{p}$ and $J/\Psi\to \rho\pi$.  It is to be
stressed that, with very few exceptions 
\footnote{The most prominent example is the $\pi\gamma$ transition
form factor. The process is special in so far as the handbag and the
leading-twist factorization fall together (see Sect.\
\ref{sec:handbag}).}, the absolute magnitudes of observables calculated 
to leading-twist accuracy, are way below experiment.
The observation that a number of hard processes respect the dimensional
counting rules is not sufficient to establish the dominance of the leading-twist 
mechanism and to rule out other explanations (see, for instance, Ref.\
\cite{DFJK2}). Scaling violations due to perturbative QCD, namely the
running of $\als$ and the evolution of the \da s, have to be
observed as well. In contrast to deep inelastic lepton-nucleon
scattering there is no experimental evidence for scaling violations in
exclusive reactions.    
\section{The handbag factorization}
\label{sec:handbag}
For hard exclusive processes there is an alternative scheme, the 
handbag factorization (see Fig.\ \ref{fig:handbag}) where only one 
parton participates in the hard subprocess (e.g.\ $\gamma q\to \gamma q$ 
in Compton scattering (CS)) and the soft physics is encoded in
generalized parton distributions (GPDs). The handbag approach applies
to deep virtual exclusive scattering (e.g.\ DVCS) where the incoming
photon has a large virtuality, $Q^2$, while the squared invariant
momentum transfer, $-t$, is small. It also applies to wide-angle 
scattering (e.g.\ WACS) where $Q^2$ is small while $-t$ (and $-u$) are
large. 

Since neither the generalized parton distributions nor the distribution 
amplitudes can be calculated whithin QCD at present, it is difficult
to decide which of the factorization schemes provides an appropriate
description  of, say, WACS at $-t\simeq 10 \gev^2$. The leading-twist
factorization probably  requires larger $-t$ than the handbag one
since more details of the hadrons have to be resolved. Recent
phenomenological and theoretical developments support this
conjecture~\cite{BK}. The ultimate decision which of the factorization
schemes is appropriate at scales of the order of $10 \gev^2$ 
is to be made by experiment.

As an example of the handbag contribution let me discuss WACS
\cite{rad98,DFJK1}. One can show 
that the subprocess Mandelstam variables $\hat{s}$ and $\hat{u}$ 
approximately equal the corresponding ones for the full process,
Compton scattering off protons. The active partons, i.e.\ the ones to 
which the photons couple, are approximately on-shell, move collinear 
with their parent hadrons and carry a momentum fraction close to
unity, $x_j, x_j' \simeq 1$. Thus, like in DVCS, the physical
situation is that of a hard parton-level subprocess, 
$\gamma q\to \gamma q$, and a soft emission and reabsorption of quarks 
from the proton. The helicity amplitudes for WACS then read
\ba
{M}_{\mu'+,\,\mu +}(s,t) &=& \;2\pi\ale 
     \big[\, { T}_{\mu'+,\,\mu+}(s,t)\,(R_V(t) + R_A(t))\,\nn\\[0.5em]
&&\qquad + \;\;  { T}_{\mu'-,\,\mu-}(s,t)\,(R_V(t) - R_A(t)) \big]  
                                                          \,,\\[0.5em]
 { M}_{\mu'-,\,\mu +}(s,t) &=& \;-\pi\ale \frac{\sqrt{-t}}{m} 
         \big[\,  T_{\mu'+,\,\mu+}(s,t)\, 
         + \,  { T}_{\mu'-,\,\mu-}(s,t)\, \big] \,R_T(t)\,.\nn
\label{ampl}
\ea
$\mu,\, \mu'$ denote the helicities of the incoming and outgoing
photons, respectively. The helicities of the protons in $ { M}$ and
quarks in  the hard scattering amplitude $ T$ are labeled by their
signs. The subprocess amplitudes have been calculated to
next-to-leading order of perturbative QCD \cite{hkm}. The form 
factors $R_i$ represent $1/\bar{x}$-moments of GPDs at zero skewness.  
$R_T$ controls the proton helicity flip amplitude while the
combination $R_V+R_A$ is the response of the proton to the emission
and reabsorption of quarks with the same helicity as it and $R_V-R_A$
that one for opposite helicities. The identification of the form
factors with $1/\bar{x}$-moments of GPDs is possible because the plus
components of the proton matrix elements dominate as in DIS and
DVCS. 

In oder to make predictions for Compton scattering a model for the 
soft form factors or rather for the underlying GPDs is required.
A first attempt to parameterize the GPDs $H$ and $\widetilde{H}$  
at zero skewness reads \cite{rad98,DFJK1}
\ba
H^a(\bar{x},0;t) &=& \exp{\left[a^2 t
        \frac{1-\bar{x}}{2\bar{x}}\right]}\, q_a(\bar{x})\,,\nn\\ 
\widetilde{H}^a(\bar{x},0;t) &=& \exp{\left[a^2 t
        \frac{1-\bar{x}}{2\bar{x}}\right]}\, \Delta q_a(\bar{x})\,,
\label{gpd}
\ea
where $q(\bar{x})$ and $\Delta q(\bar{x})$ are the usual unpolarized
and polarized parton distributions in the proton. The only free
parameter is $a$, the transverse size of the proton and even it is restricted to the
range of about 0.8 to 1.2 $\gev^{-1}$ for a realistic proton. Note
that $a$ mainly refers to the lowest Fock states of the proton
which, as phenomenological experience tells us, are rather compact.
The model (\ref{gpd}) is designed for large $-t$ which, forced by the
Gaussian in (\ref{gpd}), also implies large $x$. 
The model can be motivated by overlaps of light-cone wave
functions \cite{DFJK1,DFJK3} and it may be improved in various
ways. For instance, one may treat the lowest Fock states explicitly
or take into account the evolution of the GPDs.

From the GPD $H$ one can calculate the proton's Dirac and Compton
($R_V$) form factors by taking appropriate moments
\be
F_1=\sum_q e_q \int_{-1}^1 d\bar{x}\; H^q(\bar{x},0;t)\,,\quad
R_V=\sum_q e_q^2 \int_{-1}^1 \frac{d\bar{x}}{\bar{x}}\; H^q(\bar{x},0;t)\,.
\label{formfactors}
\ee
The axial vector form factor and $R_A$ are analogously related to the GPD
$\widetilde{H}$. Evaluation of the form factors reveals that the  
scaled form factors $t^2 F_1$ and $t^2 R_i$ exhibit broad maxima which
mimick dimensional counting in a range of $-t$ from, say, $3$ to about
$20\,\gev^2$. For very large values of $-t$, well above 
$100\,\gev^2$, the form factors gradually turn into a $\propto 1/t^4$ 
behaviour; this is the region where the leading-twist contribution
takes the lead. 

The Pauli form factor, $F_2$, and its Compton analogue $R_T$ contribute
to proton helicity flip matrix elements and are related to the GPD $E$
\be
F_2=\sum_q e_q\int_{-1}^1 d\bar{x}\; E^q(\bar{x},0;t)\,,\quad
R_T=\sum_q e_q^2\int_{-1}^1 \frac{d\bar{x}}{\bar{x}}\; E^q(\bar{x},0;t)\,.
\ee
The overlap representation of $E$ \cite{DFJK3} involves components of
the proton wave functions where the parton helicities do not sum up to
the helicity of the proton. The associated form factors are therefore 
suppressed by at least $1/\sqrt{-t}$ as compared to $F_1$ and $R_{V,A}$.
An estimate of the size of $R_T$ can be obtained by simply assuming
that $R_T/R_V$ roughly behaves as its experimentally known
electromagnetic counter part $F_2/F_1$ \cite{gayou}.  

The predictions for the Dirac form factor and the Compton cross
section are in fair agreement with experiment. The approximative
$s^6$-scaling behaviour of the Compton cross section  observed 
experimentally~\cite{shupe} is related to the broad maxima the scaled
form factors exhibit. The handbag amplitudes \req{ampl} also provide
interesting predictions for polarization observables in Compton scattering
\cite{DFJK2,hkm} among them the helicity correlation $A_{LL}$ which I
already discussed in the context of inclusive reactions in Sec.\ 2.
Within the handbag approach, the correlation between the initial state
photon and proton helicities reads \cite{DFJK2,hkm} 
\be
A_{LL} \simeq \hat{A}_{LL} \frac{R_A}{R_V} \,,
\ee  
where the $\gamma q\to\gamma q$ subprocess correlation $\hat{A}_{LL}$
is given in \req{eq:allhat}. The latter is diluted by the ratio of the
form factors $R_A$ and $R_V$ (as well as by other corrections) but its
shape essentially remains unchanged. The predictions for $A_{LL}$ from 
the leading-twist approach drastically differ from the handbag ones. 
For $\theta \lsim 110^\circ$ negative values for $A_{LL}$ are found for 
all but one examples of distribution amplitudes~\cite{dixon}. 
The JLab E99-114 collaboration \cite{nathan} has reported a first, yet
preliminary measurement of $A_{LL}$ at a c.m.s. scattering angle of 
$120^\circ$ which seems to be in agreement with the prediction 
from the handbag  while the leading-twist calculations fail badly. A
measurement of the angular dependence of $A_{LL}$ would be highly
welcome for establishing the handbag approach.
\begin{figure}[t]
\begin{center}
\includegraphics[width=5.7cm,bbllx=33pt,bblly=47pt,bburx=420pt, 
bbury=295pt,clip=true]{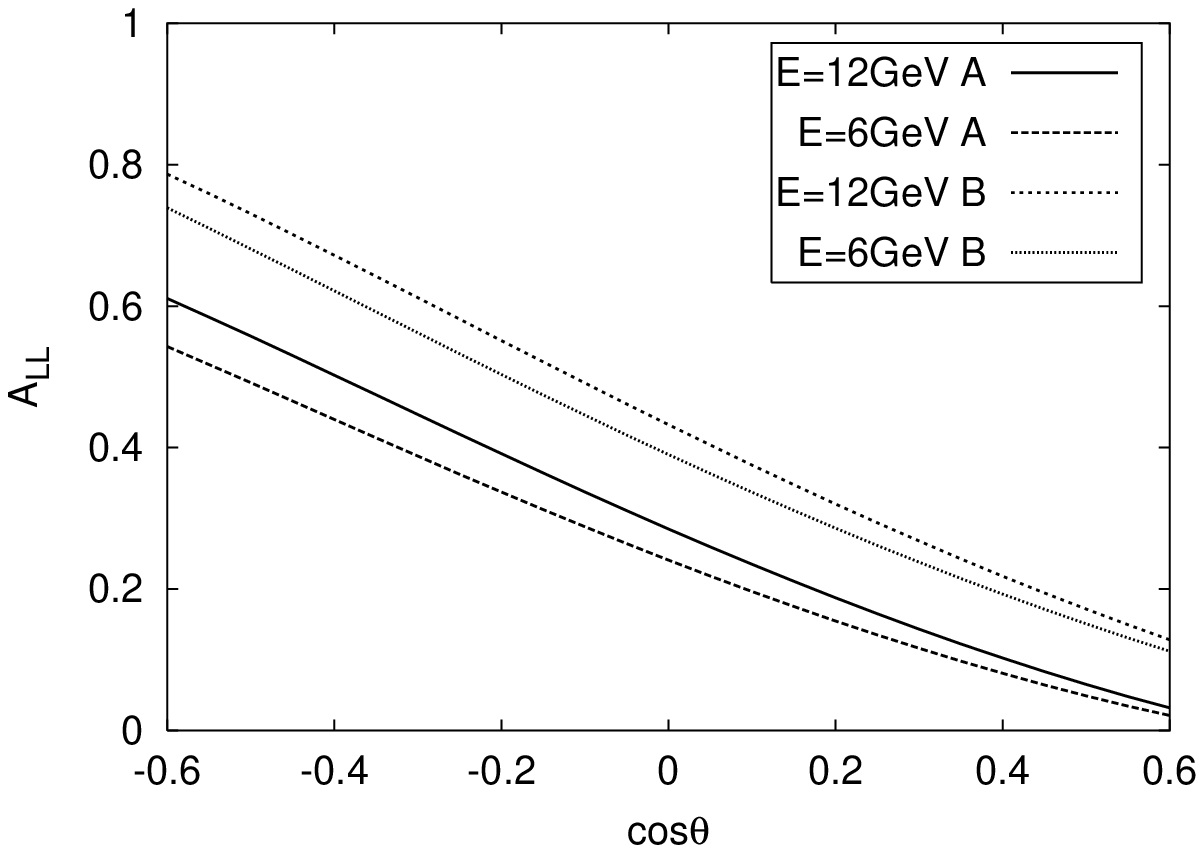} 
\includegraphics[width=5.2cm,bbllx=53pt,bblly=213pt,bburx=545pt, 
bbury=570pt,clip=true]{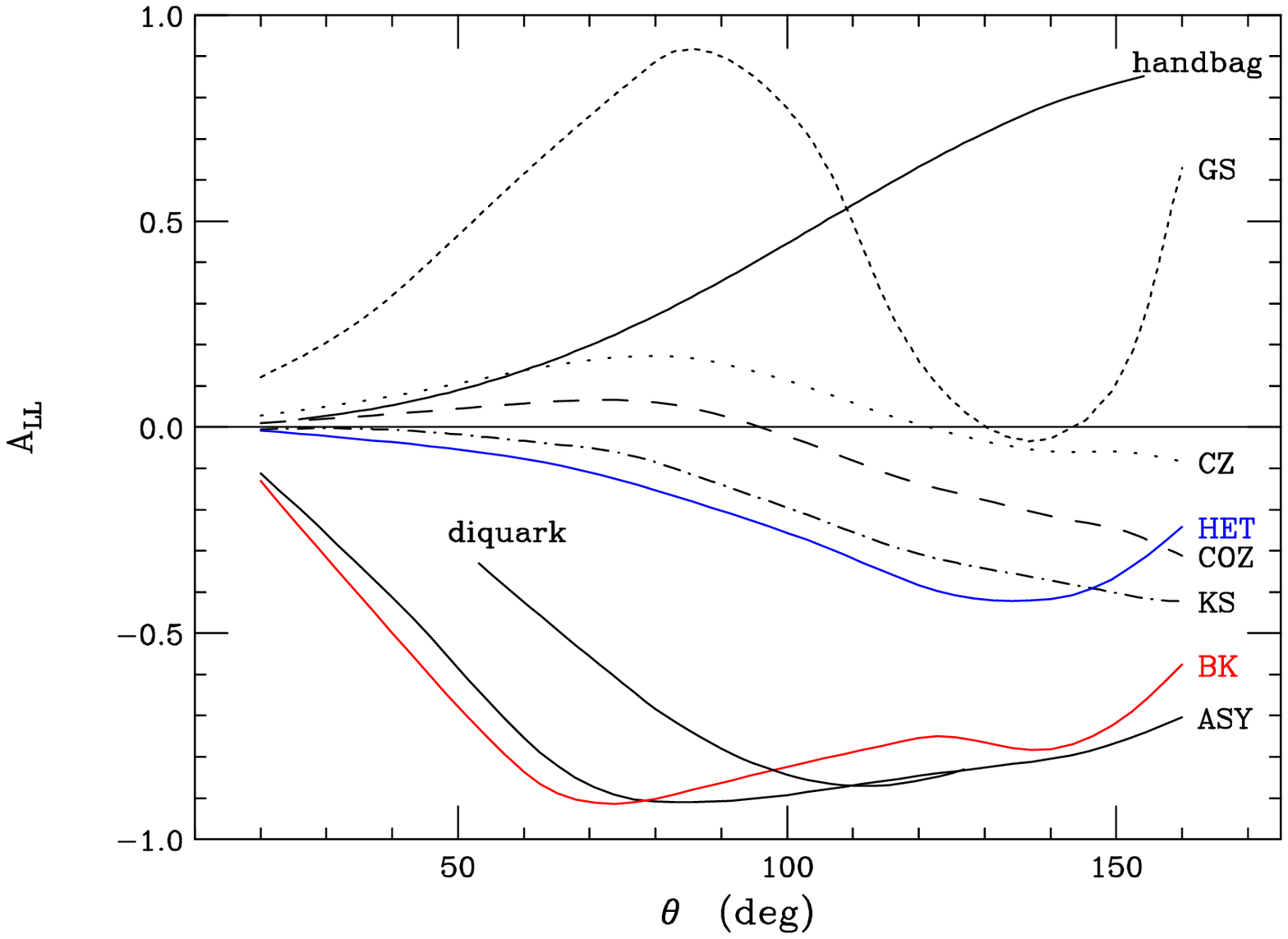}
\caption{Predictions for the helicity correlation $A_{LL}$ from the
handbag approach \protect\cite{hkm} (left) and from the leading-twist
one \protect\cite{dixon} for various \da{}s (right).}
\label{fig:cross}
\end{center}
\end{figure}

The handbag mechanism also applies to wide-angle photo- and
electroproduction of mesons \cite{hanwen}. It turns out that,
for the production of pseudoscalar mesons, $P$, the $\gamma q\to Pq$
subprocess helicity correlation coincides with \req{eq:allhat}. 
Therefore, $A_{LL}$ for the full process is very similar to that of
Compton scattering. It is, however, to be stressed that the normalization  
of the photoproduction cross section is not yet understood. 
\section{Fermion polarizations}
The polarization of the proton in two-body reactions is notoriously
difficult to calculate within QCD. It requires proton helicity flip 
and phase differences between flip and non-flip amplitudes. Both the 
ingredients are, in general, difficult to produce. Despite of this the proton
polarization in hard processes is often substantial, e.g.\ in
proton-proton elastic scattering~\cite{krisch}. 
As an example let me consider WACS again. In the leading-twist
approach hadronic helicity conservation forbids proton helicity flip
while phases are generated by on-shell going subprocess
propagators~\cite{farrar}. Thus, to leading-twist accuracy, the proton 
polarization is zero. In the handbag approach, on the other hand,
proton helicity flip is connected with the form factor $R_T$ and
phases appear in the subprocess to next-to-leading order of
perturbative QCD. Although non-zero, the proton polarization 
amounts only to a few percent~\cite{hkm}.

Another example of a fermion polarization is the beam asymmetry, $A_L$,
in $\vec{e}p\to ep\gamma$, which measures the imaginary part of the 
interference between the amplitudes for longitudinal and transversal
polarizations of the virtual photon~\cite{guichon96}. The combination
of Compton and Bethe-Heitler contribution leads to a characteristic 
dependence of $A_L$ on the azimuthal angle \cite{diehl97} which agrees 
with experiment \cite{HERMES01}.     
\section{Summary}
In gauge theories at short distances the helicity state of an
elementary particle (leptons, quarks) plays a fundamental role as its
other quantum numbers. The properties of gauge theories lead to
characteristic helicity asymmtries which may allow for a
discrimination between the leading-twist mechanism and power
corrections (as for instance the handbag) in exclusive processes or
between different subprocesses in inclusive ones. The helicity
correlation $A_{LL}$ is a particularly interesting observable because,
first, its corresponding subprocess correlation is large and,
secondly, it is often only  mildly affected by the soft
physics. Opposed to it is the polarization of the proton which is 
extremely sensitive to the soft physics and therefore difficult to predict.


\end{document}